\documentclass{bb}
\usepackage{url}
\usepackage{graphicx} 

\begin{document}

\title{Biography of eight astronomers and physicists}

\titlerunning{Biography of eight astronomers and physicists}

\author{Emmanuel Davoust \inst{1}
}
\institute{
IRAP, Universit\'e de Toulouse, CNRS, 14 Avenue Edouard Belin, F-31400 Toulouse, France\\
\email{edavoust@irap.omp.eu}\\
}

\date{Received 2012; accepted }

\abstract{
We provide a brief biography of seven French astronomers and physicists and of a Russian astronomer from the
19th and 20th centuries. Roger Bouigue (1920- ) was the director of Toulouse Observatory in the 1960s.
Claude-Louis Mathieu (1783-1875), a member of Bureau des Longitudes, worked in geodetics and celestial mechanics.
Claude Pouillet (1790-1868), a physics professor, was the first to measure the solar constant. Yves Rocard (1903-1992)
conducted research in many fields of physics, and founded the radioastronomy Observatory of Nan\c cay. Jean R\"osch
(1915-1999) was the director of Pic du Midi Observatory for over thirty years, and in this capacity lead it to the
forefront of research in high-resolution solar and planetary astronomy. 
Cyrille Souillart (1828-1898) had a lifelong interest in the theory of Jupiter's satellites. Jules Violle (1841- 1923) 
was a physics professor who determined the temperature of the solar photosphere. The Russian astronomer Kirill
Ogorodnikov (1900-1985) wrote a well-known manual on stellar dynamics.
}

\keywords{Biography, astronomer, physicist}

\maketitle

\section{Roger BOUIGUE}

Roger Bouigue was the director of Toulouse Observatory and a professor of astronomy at the Faculty of Science of Toulouse.  He developed the study of carbon stars and spectroscopic binaries at Toulouse Observatory.  He identified the relatively rare carbon stars with unusually strong isotopic bands of carbon as a homogeneous group and called them J-type carbon stars.

Roger Bouigue was born in Toulouse (France) on August 1, 1920. 
He was the son of the head of a workshop of ``Ponts et Chauss\'ees", the French Highways Agency. He was too young by a few months to be conscripted into the French army during WWII and served instead for eight months on ``Chantiers de Jeunesse en France" after the armistice.  He was a primary school teacher in the Pyrenees from 1940 to 1945, and obtained a Bachelor of Science degree from Toulouse University in 1945. He taught at a high-school in the Pyrenees for two years before joining Toulouse Observatory in 1947, replacing Pierre Lacroute who had left for Strasbourg.  He pursued his predecessor's research projects -- the spectroscopic study of binary and carbon stars, using the observatory's instruments and, occasionally, the 1.2meter telescope at Haute-Provence Observatory, where he obtained the spectra of carbon stars for his thesis. He also briefly conducted theoretical studies in molecular physics to determine the vibrational temperatures of the CN molecule and of the Swan bands of C2.  He is mostly acknowledged for having identified the common spectral features of a small category of carbon stars, those with unusual isotopic abundances of carbon, and grouped them into a single class, which he called J stars.  

Bouigue defended a thesis on two subjects, ``Contribution to the study of red carbon stars" and ``A new machine for measuring spectra", at the Faculty of Science of Paris University in 1953, with Andr\'e Danjon and the future Nobel Prize winner Alfred Kastler on the jury.

Bouigue was a talented lecturer, teaching astrophysics at Toulouse University starting in 1953, and he became one of the few full professors at the Faculty of Science in 1962.

Bouigue became the director of Toulouse Observatory in 1961, and, after finishing the astrometric reductions linked to the fourth catalogue of the Toulouse zone of the Carte du Ciel, organized research around the observation of spectroscopic binary stars.  However, since most directors of French observatories were former students of Ecole Normale Sup\'erieure, he felt that he was at a disadvantage running the Institute and interacting with astronomers at other institutes. In 1964 he participated in the search in South Africa for a site for the future European Southern Observatory. Jean R\"osch succeeded him as director in 1971.

In 1974 Bouigue became a scientific consultant for the French military laboratories and developed a theory of molecular collisions which enabled these laboratories to solve a problem linked to the ignition and combustion of small- and medium-size ordinance.

After his retirement in 1987, Bouigue worked for twenty years at the library of a local learned society and wrote a book entitled ``Positions and Motions of the Stars: Introduction to Astronomy with your PC, with programs in Basic on floppy disc". 

Bouigue was awarded the bronze medal of the CNRS (the French National Research Council) in 1955, was named Chevalier (knight) of the Ordre des Palmes Acad\'emiques in 1957 and Officer in 1980. He received the Pierre Guzman prize of the French Academy of Science in 1965 for his research in stellar spectrography. He also became corresponding member of the Royal Academy of Science of Belgium in 1968 and member of the New York Academy of Sciences in 1982.

\begin{figure}
\centering
\includegraphics[width=70mm]{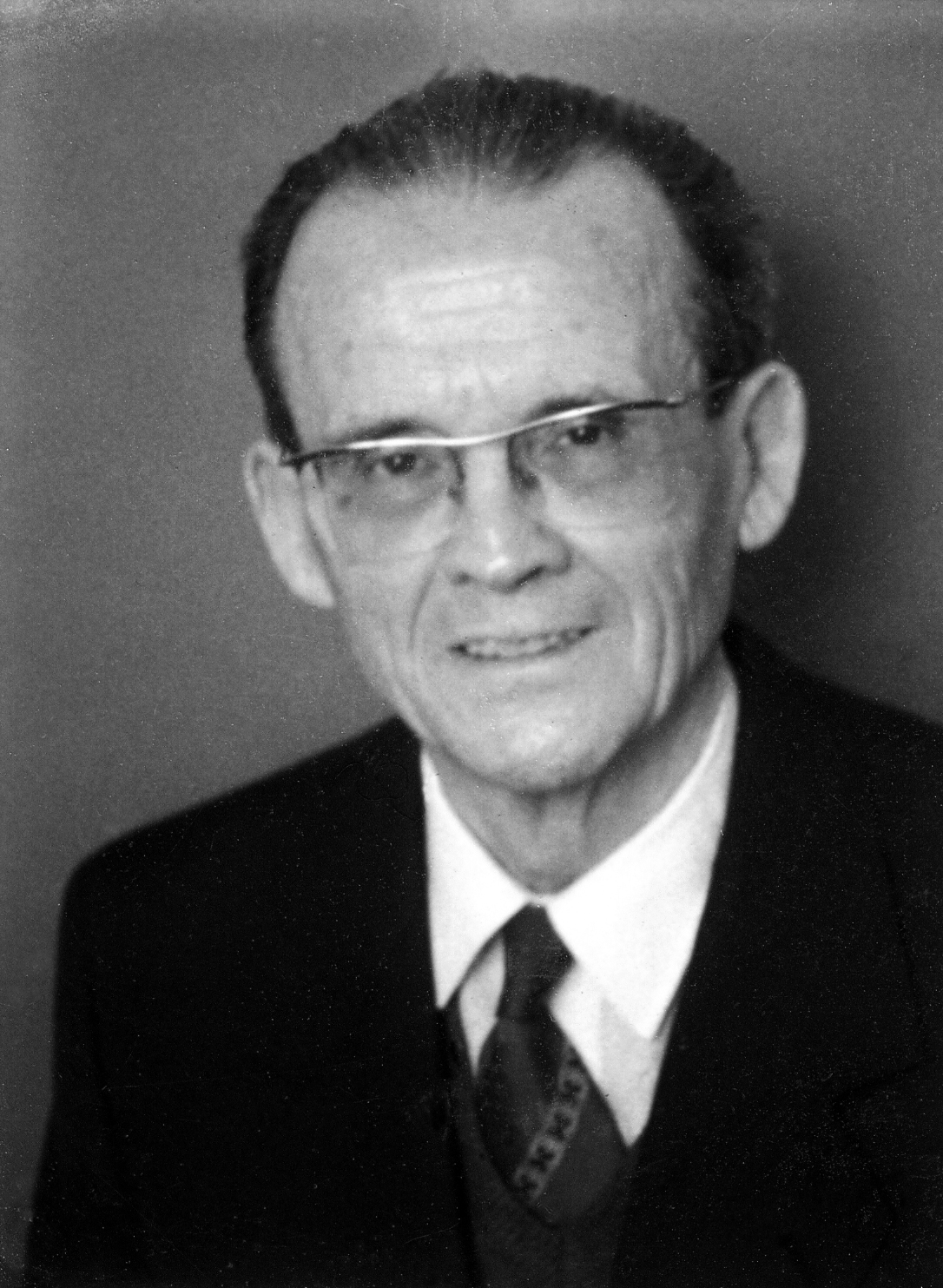}
\caption{Roger BOUIGUE in 1987}
\end{figure}

\section{Claude-Louis MATHIEU}

Claude-Louis Mathieu, who was three years older than Fran\c cois Arago, was admitted to Ecole Polytechnique the same year as Arago, and spent most of his career following in the footsteps of his friend and colleague, assisting him in various scientific endeavors. He became an active member of the scientific establishment, of the French Academy of Science, teaching at various schools, and played a significant role in French politics as a member of parliament.

Claude-Louis Mathieu was born in M\^acon (France) on November 25, 1783.
His father was a carpenter, and could not afford to pay for his education.  His early gift for mathematics were noticed by a priest in his home town, who taught him algebra and arithmetics. In 1801, he went to Paris with a letter of introduction to Joseph Delambre, who took him under his protection and housed him in his observatory. After two years of study, Mathieu passed the exam of Ecole Polytechnique where he met and became friends with Fran\c cois Arago.  

He then went to Ecole des Ponts et Chauss\'ees but quit after one year because, thanks to Delambre, he was offered the position of secretary-assistant at the Bureau des Longitudes during the absence of its incumbent.  The latter, Fran\c cois Arago, went with Jean-Baptiste Biot to the Balearic Islands to extend the measurement of the French meridian.  Mathieu participated in the routine meridian observations of stars.  Upon Arago's return in 1809, Mathieu obtained a position of astronome-adjoint at Paris Observatory in 1809, where he shared a lodging with Alexander von Humboldt.  He reduced the geodetic observations together with Biot and Johann Karl Burkhardt.  Two mutually compensating errors were later found in the reduction, and Mathieu participated in its reanalysis in 1841 with two other members of the Bureau des Longitudes, Charles-Louis Largeteau and Pierre Daussy.

Another geodetic campain in which he participated with Arago was the measurement of the difference in longitude between Paris and Greenwich in 1821-23. The measurements on the French side were never given to the British who had to rely on the published coordinates of Calais to derive the longitude difference of 9m 21.18s. A more accurate value of 9m 20.63s was obtained in 1854. 

In 1812-13 Mathieu and Arago observed the star 61 Cygni to determine its parallax.  Instead of measuring its position with respect to nearby stars, they determined the annual variations of the zenith angle of the star at the meridian. But such a method was bound to fail for parallaxes smaller than half a second of arc because of atmospheric refraction and other sources of errors.  They repeated the measurements in 1823 and correctly concluded that the parallax of the star was less than half a second.  In 1838 Bessel used the classical method to derive a parallax of 0.31, which astounded astronomers at the time at the large distance of the star.

In 1819 he participated in the observations of a comet, and in 1835 of Halley's comet, from which Arago concluded that the luminosity of comets comes from reflected sunlight.

At the request of Arago, who was then a member of the commission of lighthouses, Mathieu participated with Augustin Fresnel in experiments to improve lighthouses in 1819.  Fresnel's invention in 1820 of the lens that bears his name definitively settled the matter, to the great satisfaction of sailors.

In 1822 the Bureau des Longitudes decided to re-measure the speed of sound, considering that the value obtained in 1738 was of poor accuracy. Mathieu participated in the new measurements but the resulting value of 331.2m/s (at 0 degree centigrade) was a only slight improvement.

After the death of Delambre in 1822, leaving an unfinished History of Astronomy, Mathieu proceeded to write the missing parts, mainly about the 18th century, and the work was published in 1827.  When Arago's complete works were published by his sons a year after his death, a quarrel arose between them and Mathieu about the title of the works, and it created an enduring rift between them which resurfaced at the publication of Arago's Popular Astronomy in 1868.

In 1851 he made observations of scintillation from stars to determine if this scintillation was more important near the horizon. Nothing came of these observations, presumably because Le Verrier chased Mathieu and Laugier from their observatory position and lodging when he became the director of Paris Observatory after the death of Arago in 1853.

For most of his career, Mathieu was in charge of the publication of Connaissances des temps and of the Annuaire du Bureau des Longitudes, only surrendering the first task to Laugier late in his life.

He married Marguerite Arago, the sister of Fran\c cois Arago, in 1821.  They had a son Charles, who also went to Ecole Polytechnique, and a daughter Lucie, who married astronomer Ernest Laugier and became Arago's secretary when he was nearly blind. Mathieu and his wife also raised François Arago's three children after the death of his wife Lucie Arago in 1829.

Mathieu was active in politics, like his brother-in-law, and represented his home town from 1834 to 1849. He contributed to the law on the adoption of the metric system in France.

Mathieu received the Lalande prize of the French Academy of Science in 1809 and 1812.  He was a Commander of the Legion of Honour. He served on the jury of the London International Exhibition in 1851 and 1862, and that of Paris in 1855. He died in Paris (France) on March 5, 1875.  A bust of Mathieu was inaugurated in a public park in his home town of M\^acon in 1936.

\begin{figure}
\centering
\includegraphics[width=70mm]{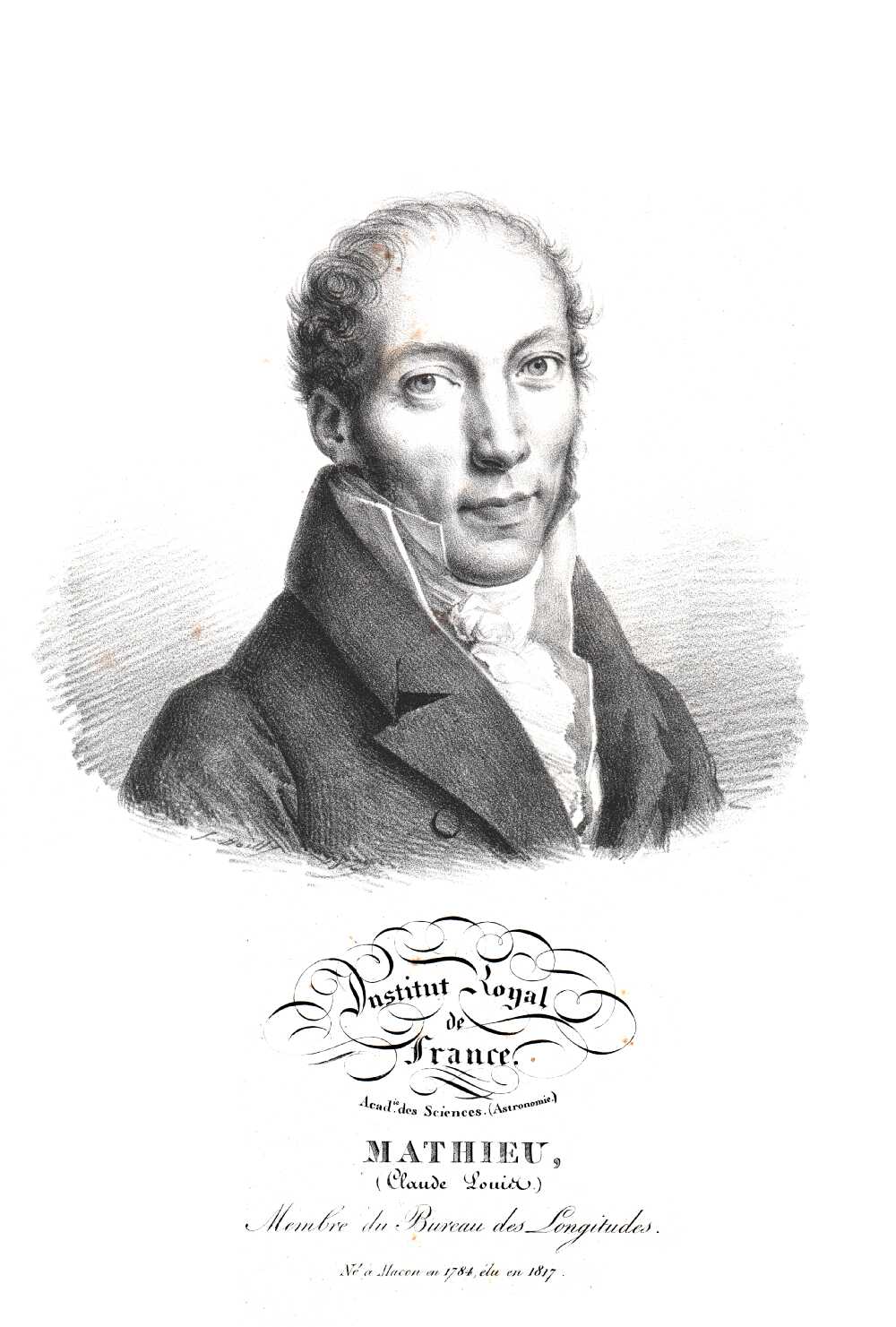}
\caption{Claude MATHIEU}
\end{figure}

\section{Claude Servais Mathias POUILLET}

Claude Servais Mathias Pouillet, educated at Ecole Normale Sup\'erieure, was a professor of physics at the Faculty of Science of Paris, a member of the French Academy of Science and the Director of the Conservatoire des Arts et M\'etiers.  He invented a pyrheliometer and was the first to measure the solar constant.  The high temperature of the solar surface that he derived drew scientists' attention  to the considerable energy radiated by the Sun and to the question of its origin.

Claude Pouillet, the son of a stationer, was born in Cusance (France) on February 16, 1790. He grew up in Besan\c con, studied at Ecole Normale Sup\'erieure between 1811 and 1813, and taught at that school until it was closed in 1822. He also taught physics at Coll\`ege Royal Bourbon, the Faculty of Science of Paris University and Conservatoire Royal des Arts et M\'etiers. He was also the director of the latter institute between 1832 and 1849.  He became full professor of physics at Paris University in 1838.  He was a talented lecturer and published a manual of experimental physics and meteorology which had six successive updated editions between 1827 and 1856. He lost his son and daughter in rapid succession shortly after retiring, which considerably saddened the end of his life.

Pouillet's early research was on optics and electricity.  He empirically established the law of electricity that bears his name, and which is a consequence of Ohm's law.  He also worked on the calorific properties of solids, and in 1822 discovered that heat was produced when a liquid is added to a powdered solid, an effect that bears his name.  In 1836 he measured the specific heat of platinum at high temperatures and the melting point of various metals and alloys with two kinds of pyrometers.  

A year later, he was the first to measure the solar constant and gave a lower limit to the temperature at the surface of the Sun. For this purpose, he invented a pyrheliometer composed of a thin cylindrical silver tube filled with water. The disk-shaped top of the tube which received the solar radiation was blackened with soot. The instrument was successively placed in the shade and in the Sun and the temperature variations recorded.  This is the dynamic method for determining the solar constant.  The detailed procedure and subsequent treatment allowed him to take into account, to first order, the heat exchanges with the environment during the experiment.  He repeated the experiments for several years, and obtained an empirical formula that linked the temperature variation $dt$ to the thickness $e$ of the atmosphere as $dt = Ap^e$, where $A$ and $p$ are constants.  He finally obtained a value of 1.7633 calories/minute/cm$^2$ for the solar constant, which is 10\% lower than the presently accepted value.

\begin{figure}
\centering
\includegraphics[width=70mm]{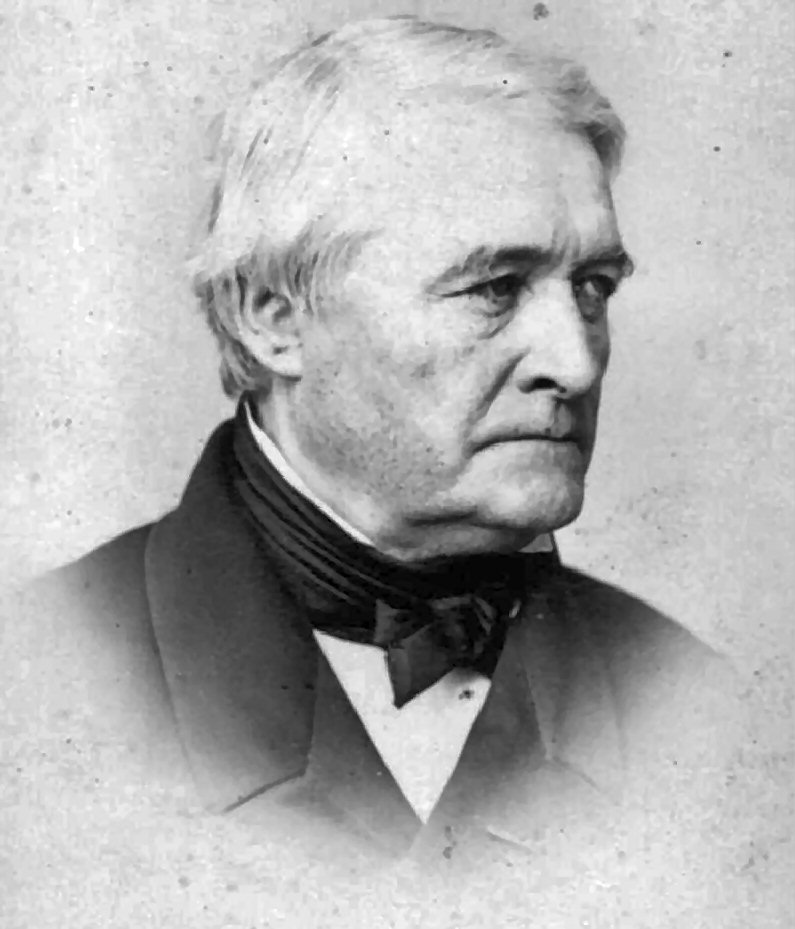}
\caption{Claude POUILLET}
\end{figure}

He used the solar constant to estimate the temperature at the surface of the Sun and found 1461 degrees, which he considered a lower limit.  If, instead of using the formula of Dulong and Petit for the energy radiated by a blackbody, he had used the Stefan-Boltzman law, which dates from 1879-84, he would have obtained 5700 K, fairly close to the current value.

Like several of his contemporaries, Claude Mathieu and Fran\c cois Arago, Pouillet was also active in politics and was a representative for Poligny (near Switzerland) at the parliament between 1837 and 1848.

Claude Pouillet was elected to the French Academy of Science in 1837. He died in Paris (France) on June 14, 1868.

\section{Yves ROCARD}

Yves Rocard, educated at Ecole Normale Sup\'erieure, was a very creative physicist,  who conducted research in various fields: the diffusion of light in fluids, electronics, the propagation of electromagnetic waves, radio astronomy, and atomic physics. 

Yves Rocard was born in Vannes (France) on May 22, 1903. He came from a modest background. His father, a captain in the French Air Force, died when his plane was shot down by the Germans during WWI. He was admitted to Ecole Normale Sup\'erieure in 1922. His first research project as a graduate student was to study the diffusion of light in liquids, under the supervision of Charles Fabry in 1927. His results were not well received by the jury because he found about 50\% more diffused light than predicted. Only later did he understand that he had almost discovered the Raman effect, which earned Raman the Nobel prize in 1930.  He then worked as an engineer in a company producing lamps for radios.  In 1939 he was named assistant professor at Paris University.

During WWII, he worked for the National Defence and joined the French resistance after the armistice. In 1943 he volunteered to determine the principle and method of operation of the radio beacons (called Bernhardiner) installed by the Germans on the French west coast and reported his findings to the British Intelligence Service.  Not being able to return to France, he then designed radars for the ships of the Free French Naval Forces.

In 1945 he was named director of his former alma mater in replacement of Georges Bruhat, who had died in the concentration camp of Sachsenhausen, and reorganized teaching and research in physics. He remained in charge of research for the French Navy and in 1946 initiated the service of ionospheric forecasts, which used the coronographic observations made at Pic du Midi Observatory.

In 1946 he started a research team on radio astronomy to study solar radio emission, because he knew from experience that airplanes disappeared from radar screens when flying in front of the Sun.  The team, first composed of Jean-Fran\c cois Denisse and Jean-Louis Steinberg, adapted surplus radar antennas and receivers from WWII for their purposes.  Most of the research was performed with two 7.5-meter ``W\"urzburg Riese" antennas at a Navy research laboratory in Marcoussis near Paris.

As the radio background noise was important in the Parisian region, Rocard obtained important funds from the Ministry of Education in 1952 to build a radio astronomy station at Nan\c cay, 200 km south of the capital. The research team moved to Paris Observatory in Meudon thereafter, and the station was inaugurated in 1956.

Rocard also created a research team in solid state physics at Ecole Normale Sup\'erieure, which would later favor the development of millimeter-wave radio astronomy at that school.  

He later became a consultant for the French Atomic Energy Commission which developed the French atomic bomb. His late research on biomagnetism and his gift as a water diviner earned him the wrath of the scientific establishment and cost him a seat at the French Academy of Science.

His son Michel Rocard became prime minister during the presidency of Fran\c cois Mitterand.

Yves Rocard was awarded the Holweck prize and medal by the British Physical Society and the Soci\'et\'e Francaise de Physique in 1948.  He was awarded the Henri de Parville prize in 1942 and the Lamb prize (with two other scientists) in 1960  by the French Academy of Science. He died in Paris (France) on March 16, 1992.  The Soci\'et\'e Francaise de Physique created an annual prize in his name in 1992, as a tribute to his work in physics.

\section{Jean R\"OSCH} 

For over thirty years Jean R\"osch was the director of Pic du Midi Observatory, which he brought to the forefront of international research in solar astrophysics, planetary and lunar astronomy and cosmic ray physics. His own research was aimed at the effects of the atmosphere on the quality of images and at ways to improve the latter. He was also a professor of astrophysics at Pierre et Marie Curie University of Paris and an active member of national and international scientific institutions.

Jean R\"osch was born on January 5, 1915 in the Algerian garrison town of Sidi-bel-Abb\`es where his father, of Danish origin, was a medical doctor in the army.
He went to high school in Algiers.  He joined the French Astronomical Society and visited the Pic du Midi Observatory in 1931 and again in 1936, during an internship with Bernard Lyot. He was accepted at Ecole Polytechnique and Ecole Normale Sup\'erieure in 1933, and chose the latter. He married Raymonde Postel, who was also from Sidi-bel-Abb\`es, in 1937.  The couple remained childless. That same year he was conscripted as an artillery officer and did not return to civilian life until after the armistice of June 1940. In 1939 he joined a group of scientists studying new techniques, radar  
in particular, for detecting aircraft.  He occasionally pursued this research after the war in his capacity as a reserve officer.

\begin{figure}
\centering
\includegraphics[width=70mm]{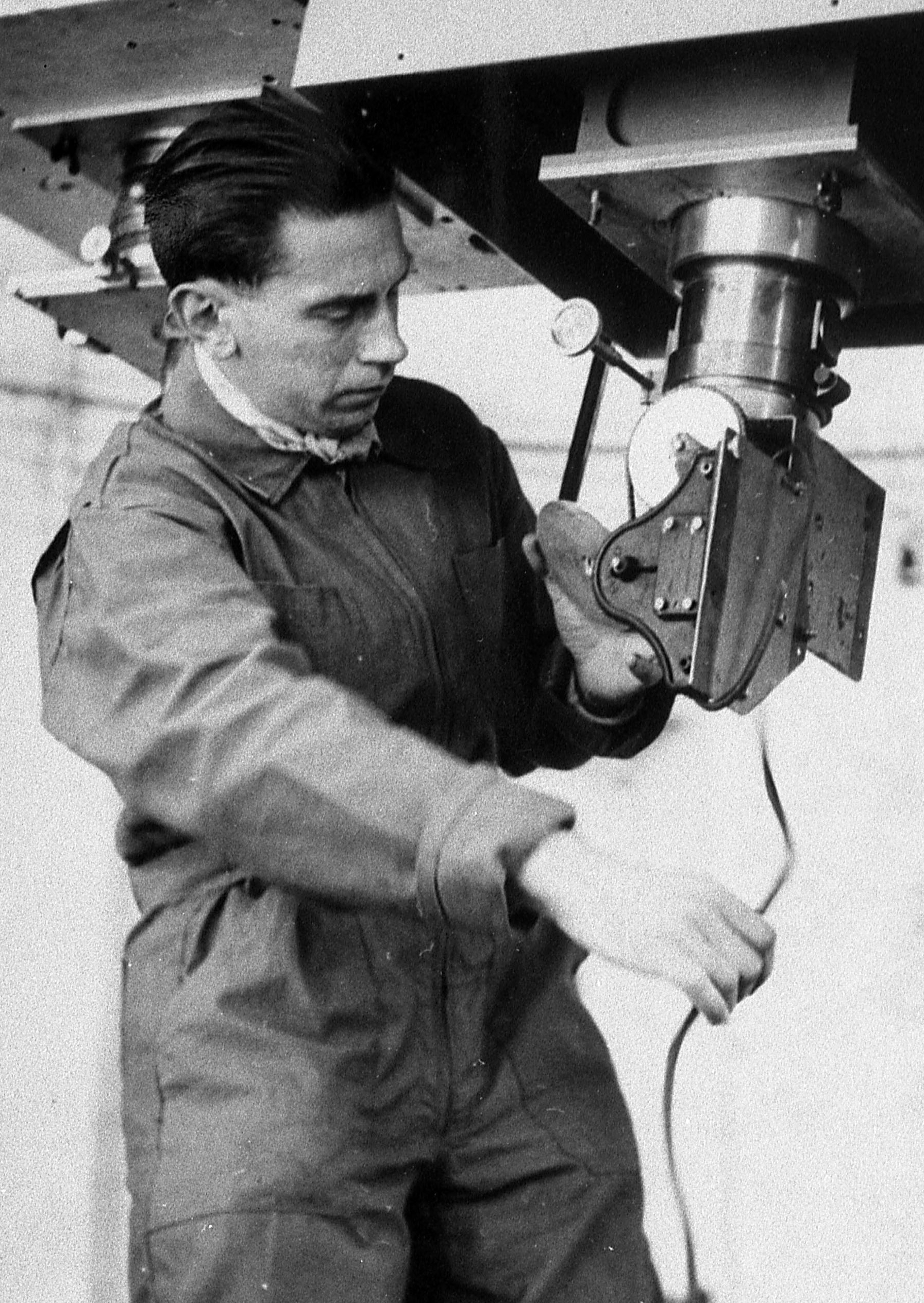}
\caption{Jean R\"osch at the Baillaud refracto-reflector}
\end{figure}

In 1940 he joined Bordeaux Observatory where he investigated the possibility of measuring astronomical distances by stereoscopic methods akin to the ones he had studied in the army and defended in a thesis on that subject in 1943.

When he was named director of Pic du Midi Observatory in 1947, the situation at this high altitude observatory was, in his own words, one of hopes rather than achievements.  His predecessor, Jules Baillaud, had worked unremittingly for ten years to modernize the observatory, despite his age and the catastrophic conditions of wartime, and most projects had only made progress on paper.

The first five years of his directorship were devoted to the task of improving the infrastructure. The electric line reached the summit in November 1949, finally allowing the cosmic ray physicists to start their electromagnet, and the staff to use electrical appliances without restraint.  The first regular passenger of the cable car reached the summit in August 1952, marking the end of seventy years of isolation, and of long, tedious and sometimes dangerous climbs to reach the observatory. Living quarters were limited on this narrow peak, and R\"osch's next task was to have it excavated to provide more space for rooms, laboratories and storage.

Taking advantage of the excellent transparency and quietness of the upper atmosphere, R\"osch started observing the solar photosphere in 1953, and the images obtained at Pic du Midi eventually became the finest in existence, rivaling the ones obtained by Martin Schwarzschild with his stratoscope experiment, but much cheaper as R\"osch would point out.  The first images were obtained with a 23cm refractor cooled by a flow of water of seven liters/minute, then with the 38cm lens of Toulouse Observatory made by the Henry brothers, mounted in 1959 on a dedicated telescope with a turret-shaped dome, replaced by a 50cm lens after 1971.  The reputation of the site encouraged solar astronomers from Paris Observatory to install a coelostat and two spectrographs at the summit for the study of prominences in the late 1950s.

The initial success of the photospheric observations, as well as the well-established leadership of the coronographic observations set up by Bernard Lyot at the Pic du Midi after the war, led R\"osch to a policy of promoting {\it all} astronomical observations requiring high resolution, and {\it only} those.  To that end, he had new instruments constructed and installed, such as an electronographic camera in 1962, a one-meter telescope in 1964, a coronometer in 1967, a polarimeter for prominences in 1974, a two-meter telescope in 1981, a coronograph-spectrograph in 1984. 
 
His personal research focused on the study of the perturbing influence of the atmosphere and of methods to compensate this negative effect. He was among the first to observe the pattern of speckles appearing on short exposures of tight binary stars with an electronographic camera.  Unfortunately, he did not take the time to understand and interpret these patterns; this was done in the 1970s by Antoine Labeyrie.  R\"osch tried to improve the seeing at the Baillaud refracto-reflector, by insulating its tube and blowing a current of cold air into it. He also tried to separate the images produced by the 60 cm lens of the Baillaud telescope into seven elementary images, thereby inventing pupil segmentation.  He designed domes adapted to the site, like the turret-dome where the telescope tube is inside another tube protruding from the dome.  For the dome of the two-meter telescope, he invented a complex system whereby the telescope tube (on an equatorial mounting) always had to be in front of its circular aperture, which was offcenter on a disk mounted on the side of the dome, which itself rotated in azimuth, requiring sophisticated piloting software.

He also occasionally did some mainstream astronomy, observing the transits of Mercury in 1960 and 1970 to measure its diameter accurately, and the long solar eclipse of 30 June 1973 in Mauritania.

His final research project, which started in 1971, was to build a scanning heliometer designed to measure the oblateness of the solar disk. This measurement is of importance to verify Einstein's theory of General Relativity.  The first measurements were obtained in 1984, but R\"osch worked unremittingly (as always) with a constant stream of interns to improve the instrument until shortly before his death.

In 1963, he was named professor of astrophysics at Pierre et Marie Curie University of Paris, replacing Andr\'e Danjon who retired.  This implied not only an extra workload, but also ceaseless travels by night-train between Bagn\`eres and Paris. But the advantage was being in direct contact with the future generations of astronomers, enticing some of them to devote their career to the Pic du Midi. Incidentally, he was the president of my thesis jury in 1977.

Jean R\"osch played an active role in many local, national and international scientific organizations, such as the CNRS, the French Space Agency, the French National Committee for Astronomy, the International Astronomical Union, the European Space Research Organisation, the Committee on Space Research (COSPAR).  At the IAU, he was the president of Commission 9 (astronomical instruments), of the working group on astronomical site selection (1964-1967) and of the commission for the study of the site of the European Southern Observatory (1964). He was also a member of seven learned societies, and presided four of them.

Jean R\"osch had a plaque engraved and set at the entrance of the Observatory, which read : ``Greatness first - and always - arises from a goal outside oneself.  Antoine de Saint-Exup\'ery".  For him, the goal was the Pic du Midi, and the means was a deep love of the mountain and of science. He was a gifted scientist, lucid, hard working, with a very good memory, loyal to his colleagues and unselfishly devoted to his goal, to the extent of antagonizing the Parisian establishment, who, to his one regret, did not let him join the Academy of Science.

Jean R\"osch was awarded the Benjamin Valz prize in 1942 for his research in physical astronomy and the Janssen medal in 1944 for his research in astronomical stereoscopy by the French Academy of Science, the d'Arsonval prize in 1943 for his research on the physiology of perception by the General Psychological Institute of Paris, and the Manley-Bendall medal by the Academy of Science, literature and arts of Bordeaux in 1966. He was named Chevalier (knight) of the Legion of Honour and Officer of the Palmes Acad\'emiques. He died in Bagn\`eres-de-Bigorre (France) on January 20, 1999.

\section{Cyrille Joseph SOUILLART}

Cyrille Souillart was a professor of mechanics and astronomy at the University of Lille who had a lifelong interest in the theory of Jupiter's satellites.

Cyrille Souillart, the son of a primary-school teacher, was born in Bruay-en-Artois (France) on january 20, 1828.  In 1851 he was accepted at Ecole Normale Sup\'erieure, where Victor Puiseux's lectures sparked his interest for celestial mechanics. He was a high-school teacher from 1854 to 1874.  He defended a thesis entitled ``Analytical theory of Jupiter's satellites" at Paris University in 1865.  In 1873 he was named professor of rational and applied mechanics, and in 1887 full professor of astronomy, at the University of Lille, where he was a colleague of Joseph Boussinesq.

The theory of motion for Jupiter's satellites was first developed by Pierre-Simon de Laplace in Book eight of his Treatise of Celestial Mechanics. Laplace determined an approximate analytical solution for this motion by perturbation theory.  Souillart's lifelong work was to improve on this solution, essentially by expanding the perturbing force to higher order.  He was also inspired by a method that Sim\'eon Poisson developed for his theory of Lunar motion.  F\'elix Tisserand reviewed Souillart's theory in his Treatise of Celestial Mechanics in 1896.  In this form, the theory is now known as the Laplace-Souillart theory, and was used for the satellites of Saturn and Uranus in the 1970s and 1980s.

Souillart participated in the re-edition of Laplace's work in the 1890s. He earned the Lalande prize in 1882 and the Damoiseau prize of the French Academy of Science in 1886, and was named Knight (Chevalier) of the French Legion of Honour in 1891. He died in Lille (France) on May 9, 1898.

\section{Jules Louis Gabriel VIOLLE} 

Jules Violle, educated at Ecole Normale Sup\'erieure, was a physics professor at the Universities of Grenoble and Lyon and at Conservatoire des Arts et M\'etiers. He designed an actinometer with which he measured the temperature of the Sun from the top of Mont Blanc, in the Sahara and during solar eclipses.

Jules Violle was born in Langres (France) on November 16, 1841.
He was the son and the grandson of mathematicians and was himself the father of five sons.  In 1861 he was admitted to Ecole Polytechnique and to Ecole Normale Sup\'erieure and chose the latter, where he met and befriended Edouard Branly. He then taught at several high schools and then returned to Ecole Normale Sup\'erieure at the invitation of Louis Pasteur. In 1870 he defended a thesis composed of two subjects, the mechanical equivalent of heat and the work of Pasteur on fermentation.

In 1872 he was given a faculty position at Grenoble University, where he began research on the solar radiation.  To measure the temperature of the Sun, Violle used an actinometer invented by agronomer Adrien-Etienne de Gasparin.  The temperature of the thermometer placed in a cryogenic enclosure  (a copper sphere blackened with soot and filled with ice) is measured before, during and after opening a small aperture in the enclosure to allow the Sun to shine on the thermometer.  The ingenuity of Violle was to use several apertures and thermometers, in order to determine the cooling due to the air and the heating due to the radiation coming from the sky around the Sun, thus enabling him to make absolute rather than relative measurements. 

To estimate the effect of the atmosphere, one method was to make simultaneous measurements at different altitudes. In 1875, Violle transported his actinometer to the top of Mont Blanc, while his colleague used an identical instrument at the foot of the mountain. Another method was to make successive measurements at a given location during the day, when the sunlight crosses different thicknesses of atmosphere. However, the water vapor content of the atmosphere also varies during the day, so Violle chose to resume his experiments in the Sahara desert, where the air is very dry and the water vapor presumably constant.  He estimated that the solar constant was 2.54 calories/minute/cm$^2$ and the temperature of the Sun was about 2500 degrees. Chemist Marcellin Berthelot argued that its temperature must be higher since, properly focussed, the solar radiation was able to make platinum melt. Violle answered that the temperature certainly varied accross the surface of the Sun and might be higher at some point on the surface.

While Violle's estimates of the temperature at the surface of the Sun was of considerable interest to astronomers, his own motivation was meteorology and its impact on agriculture, and his audience was primarily meteorologists.  He wanted to quantify the effect of absorption by the atmosphere on the intensity of solar radiation and to estimate in this way the quantity of water vapor in the atmosphere.

He then moved to other fields of research: he defined a photometric unit independent of the properties of the light emitter (later called the Violle standard), invented a calorimeter, investigated the velocity of sound, the effect of electric lines on hail storms, determined the specific heat and melting point of platinum. 

He remained an authority on the measurement of the solar radiation as he gave reviews on the subject at the international meteorological congresses of Rome in 1879 and of Saint-Petersburg in 1899, and in 1913 was chairman of the committee on the measurement of solar radiation of the International Union for Cooperation in Solar Research.

He returned to measuring solar radiation on the occasions of the total solar eclipses of May 1900, August 1905 and April 1912.  The instrument-maker Jules Richard designed for him a recording actinometer small and light enough to be airborne by balloon, which enabled him to obtain measurements recorded at high altitudes during the eclipses.
 
Jules Violle was a good lecturer and wrote a physics manual that was praised for its clarity.  He was elected to the French Academy of Science in 1897 and to that of Agriculture in 1909. He died in Fixin (France) on September 12, 1923.

\vfill
\section{Kirill Fedorovich OGORODNIKOV}

Kirill Ogorodnikov was a professor of astrophysics and a research scientist at Leningrad State University.  His field of research was the kinematics and dynamics of star clusters and galaxies and his textbook on stellar dynamics became a worldwide reference on the subject. 

Kirill Ogorodnikov was born in Pavlovsk (Russia) on July 30, 1900. His father was a lieutenant-general in the tsar's army and a professor of statistics and geography at the Academy of the General Staff. His paternal grandfather was an ethnographer and statistician.  Both Kirill and his father joined the Red Army after the revolution of 1917. In 1920 he was admitted to the militarised Tukhachevskij Polytechnical Institute of the western front in Smolensk and a year later to the Faculty of Physics and Mathematics of Moscow University. There, he was a member of Lusitania, a group of young mathematicians working under the direction of the future academician Nikolaj Lusin.  In 1922, while still a student, he started working at the newly founded Astrophysical State Institute (later Shternberg State Institute) and studied the motion of the Sun in the Galaxy.  After graduating from university in 1924, he went to the Institute of Mathematics and Mechanics of Moscow State University where his dissertation on the treatment of discordant observations earned him the degree of Candidate of Science in 1929. He became a professor of astronomy and geodetics in 1931 and a doctor of physical and mathematical sciences in 1936, without having defended a thesis.

Ogorodnikov was sent to Harvard at the end of 1930 and stayed there for over a year, studying the systematic motion of stars in the Galaxy. In a letter to B. Gerasimovich in 1931, Harold Shapley described Ogorodnikov as ``so instinctively mathematical that he has difficulty in seeing the full implications of a problem, and he has a tenacity of view that some people found rather unshakeable." 

Cosmology and Einstein's Theory of Relativity became an active and controversial field of research in the 1930s. Ogorodnikov joined the debate and, together with Soviet ideologist Vartan Ter-Oganezov, criticized the expanding universe model in favor of a dialectical materialist cosmology.  He considered that the expanding system of galaxies was a local phenomenon and defended the model developed at the turn of the century by Carl Charlier, where stellar systems are organized in a hierarchy of systems of increasing dimensions.

In 1932 Ogorodnikov published a model for the motion of B-stars in the Galaxy, which extended Oort's constants to rotation, shear and dilation in a general flow field.  Known as the Ogorodnikov-Milne model, because a similar model was independently proposed by Edward Milne in 1935, it is still currently used for studying the global kinematics of stars in the Galaxy.  Later in the decade he proposed a method to determine the extinction in interstellar dark clouds and their distance from star counts (later known in Russia as Ogorodnikov's method). 

In 1934 Ogorodnikov joined the staff of Pulkovo Observatory. He was named professor in 1940, and full professor in 1946, at Leningrad State University.  At the outbreak of war he volunteered in an infantry unit with the rank of captain and defended the Pulkovo heights.  He also translated propaganda texts, some of which were his own, into German and read them in German.  During such a broadcast in 1942 he was shell-shocked and hospitalized.  He was discharged from the army for health reasons and became a supervisor at the faculty of mathematics and mechanics of Leningrad University, which had been evacuated to Saratov.  He was the dean of that faculty until 1948.

He made the first review in Russian on the dynamics of rotating stellar systems in 1948 and wrote a monograph on the dynamics of stellar systems in 1957, which was published in English in 1965 and thereafter often used and cited worldwide. The originality of this work was to apply statistical mechanics and the theory of rotating fluid bodies to galaxies in order to determine their most probable state and their equilibrium properties.  In his book, he pointed out the paradox between the apparent relaxed appearance of elliptical galaxies and their long relaxation time.  After reading this book, Donald Lynden-Bell came up with the concept of violent relaxation, which solved the paradox.  Several of Ogorodnikov's PhD students, such as Tateos Agekian and Vadim Antonov, were recruited at Leningrad State University and made significant contributions to the field of stellar dynamics.

During his career, Ogorodnikov took on many editorial duties, on the editorial staff of Astronomicheskij Zhurnal (1932-37), of Zemlya i Vselennaya (1965-), head editor of the journal of abstracts Referativnij Zhurnal. He published four books of popular astronomy and one on the history of astronomy.

Ogorodnikov was a member of the astronomical committee of Narkompros (the agency in charge of public education and culture) from 1932 to 1972, a member of the Communist Party starting in 1940, and one of the first Russian members of the International Academy of Astronautics. He was honored Worker of Science of the RSFSR and awarded the Order of the October Revolution and the Order of the Red Banner of Labour for his numerous scientific and pedagogical contributions.  Ogorodnikov had many interests in life, he was a music lover, played the piano and also had a gift for drawing.  
He died on June 29, 1985, presumably in Leningrad (USSR).

\begin{acknowledgements}
We thank Roger Bouigue for giving us a detailed
account of his career, and the librarian of Observatoire Midi-Pyr´en´ees
for assistance in obtaining obituaries in obscure publications.
\end{acknowledgements}

\end{document}